# Magnonic band structure in vertical meander-shaped CoFeB thin films

Gianluca Gubbiotti[1*], Alexandr Sadovnikov[2], Evgeny Beginin[2], Sergey Nikitov[2,3,4], Danny Wan[5], Anshul Gupta[5], Shreya Kundu[5], Giacomo Talmelli[5,6], Robert Carpenter[5], Inge Asselberghs[5], Iuliana P. Radu[5], Christoph Adelmann[5], and Florin Ciubotaru[5]

[1] Istituto Officina dei Materiali del CNR (CNR-IOM), Sede Secondaria di Perugia, c/o Dipartimento di Fisica e Geologia, Università di Perugia, I-06123 Perugia, Italy.

[2] Laboratory Metamaterials, Saratov State University, 83 Astrakhanskaya street, Saratov 410012, Russia

[3] Kotel'nikov Institute of Radioengineering and Electronics, Russian Academy of Sciences, Moscow 125009, Russia

[4] Terahertz Spintronics Laboratory, Moscow Institute of Physics and Technology (National University), 9 Institutskii lane, Dolgoprudny, 141700, Russia

[5] Imec, 3001 Leuven, Belgium

[6] Departement Materiaalkunde, KU Leuven, 3001 Leuven, Belgium

**ABSTRACT**

The dispersion of spin waves in vertical meander-shaped CoFeB thin films consisting of segments located at 90° angles with respect to each other is investigated by Brillouin light scattering spectroscopy. We reveal the periodic character of several dispersive branches as well as alternating frequency ranges where spin waves are allowed or forbidden to propagate. Noteworthy is the presence of the frequency band gaps between each couple of successive modes only for wave numbers $k=n\pi/a$, where $n$ is an even integer number and $a$ is the size of the meander unit cell, whereas the spectra show propagating modes in the orthogonal film segments for the other wavenumbers. The micromagnetic simulations and analytical calculations allow us to understand and explain the results in terms of the mode spatial localization and symmetry. The obtained results demonstrate the wave propagation in three dimensions opening the path for multi-level magnonic architectures for signal processing.

**Corresponding author:**

* gubbiotti@iom.cnr.it

**ORCID: 0000-0002-7006-0370**

**Introduction**

The concept of magnonic crystals (MCs), magnetic metamaterials with periodically modulated properties on the nanoscale, was proposed about twenty years ago as the magnetic counterpart of photonic crystals.[1,23] MCs exhibit engineered and reprogrammable spin-wave (SW) band structures with alternating frequency gaps and allowed bands for SW propagation.[4] The MCs with periodicity in one (1D) and two-dimensional (2D) have been widely studied in various arrangements[5,6,7,8] and played a central role in the research field of magnonics which addresses the use of spin waves, or magnons (the SW quanta), as a tool for signal processing, communication and computation due to the low-energy-consumption property and potential compatibility with the next generation of circuits beyond CMOS electronics.[9]

In recent years, in order to explore new physical effects and functionalities, there has been increasing attention given to structures that are periodic in the film plane but inhomogeneous along the perpendicular direction.[10,11,12] The inhomogeneity along the out-of-plane direction can significantly influence SW propagation in the periodic landscape of the plane ferromagnetic film. For example, the vertical stacking of ferromagnetic materials offers more coupling mechanisms, i.e. interplay between dipolar and interlayer exchange due to the smallest interlayer separation whereas in 2D systems the coupling is essentially based on the long-range dipolar interaction. Single and coupled waveguides were considered only in the planar configuration, e.g. directional couplers performing the functions of a waveguide crossing element and frequency-selective magnonic drop filters in complex magnonic circuitry.[13,14,15] The interest in vertically coupled magnetic structures follows a similar trend in CMOS electronics,[16] where it enables a transition from two-dimensional to three-dimensional architecture and also provides a possible roadmap for further scaling.[17,18] The advantages of 3D over 2D magnetic systems are also related to the different areas of research including physics effects and potential applications.[19] For example, one can explore unidirectional SW propagation, indirect band gaps[20] or curvature-induced novel dynamic effects.[21] From the applications point of view, 3D integration permits more functionality to fit into a smaller space, thereby allowing a larger number of vertical interconnections between the layers, and an increased element density for the fabrication of scalable and configurable magnonic networks.

In this work, we propose MCs fabricated into nanometrically thick vertical meander-shaped CoFeB films consisting of ferromagnetic segments located at 90° angles with respect to each other, allowing for SW propagation in three dimensions without significant losses in the junction region. This permits to overcome the limitations of SW manipulation and steering[22,23] which is hard to be realized for in-plane magnetized films due to the SW anisotropic dispersion that depends on the relative orientation between the magnetization and the wave vector. The measured magnonic band structure exhibits full band gaps (BGs), whose width and center frequency depend on the CoFeB film thickness. Micromagnetic simulations accurately reproduce the experimental dispersion and the simulated mode profiles reveal the presence of extended and quantized standing spin-wave modes in the horizontal and vertical segments. Moreover, the formation of the magnonic band gaps as well as the dependence of their width and center frequency on the geometric film parameters are explained by theoretical calculations performed using the transfer matrix method (TMT).

Vertical meander-shaped ferromagnetic films can be considered as a prototype of 3D integrated magnonic structures that permit the vertical shifting of propagating SW from one layer to another placed at different heights. This opens up the prospect of increasing the density of magnonic elements for scalable networks and the realization of the multilevel architecture of SW signal processing.

**Sample fabrication**

Magnetic $Co_{40}Fe_{40}B_{20}$ (CoFeB) meander-shaped films were deposited by physical vapor deposition on periodic $Si/SiO_2$ substrate whose surface is periodically corrugated by grating with a height of $h$=50 nm and a width of 300 nm. The array lattice constant is $a$=600 nm resulting in an edge of the Brillouin zone (BZ) in reciprocal space equal to $\pi/a$=0.52x10$^7$ rad/m. Two samples with different thicknesses ($d_1$) and widths ($d_2$) of the CoFeB horizontal and vertical segments, respectively, have been fabricated. A schematic drawing of the meander-shaped film is shown in Fig. 1 (a). The geometric parameters for the first sample are given by $d_1$=23 nm and $d_2$=12 nm whereas for the second sample $d_1$=15 nm and $d_2$=8 nm. In the remaining part of the paper, these samples are labeled as $d_1$-$d_2$, i.e 23-12 and 15-8 samples, respectively. Scanning electron microscopy (SEM) images of the 23-12 meander-shaped CoFeB film are represented in Fig. 1 (b) and (c). It can clearly be seen that the CoFeB film conformally coats the Si grating, as well as well-defined vertical segments of nearly 90° angles.

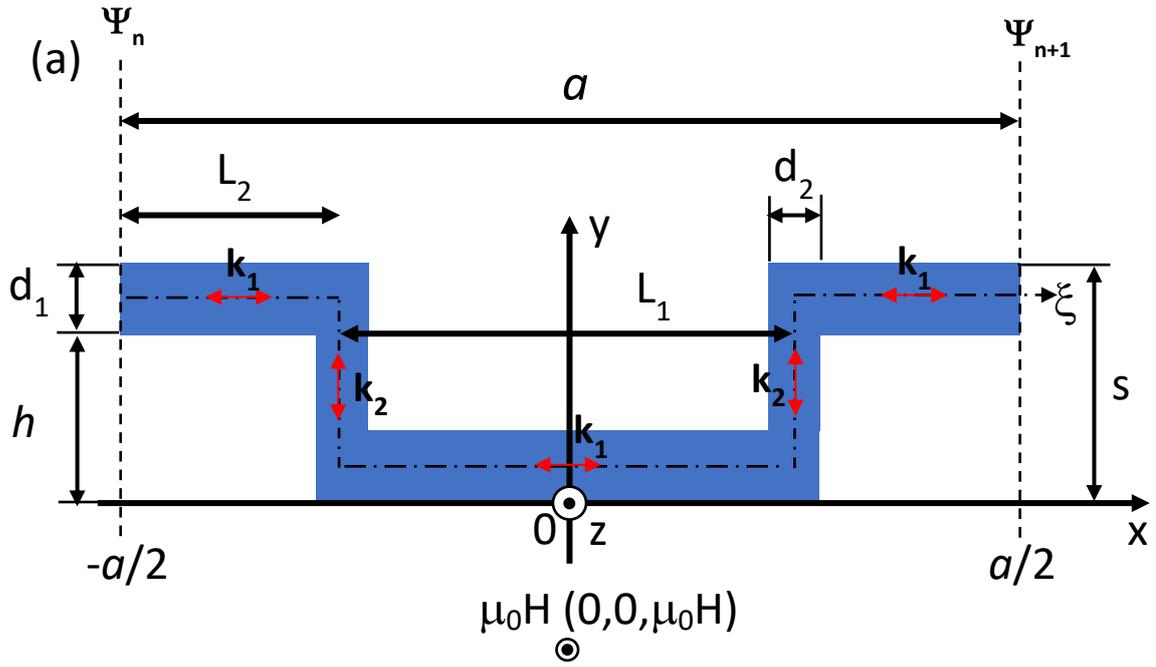
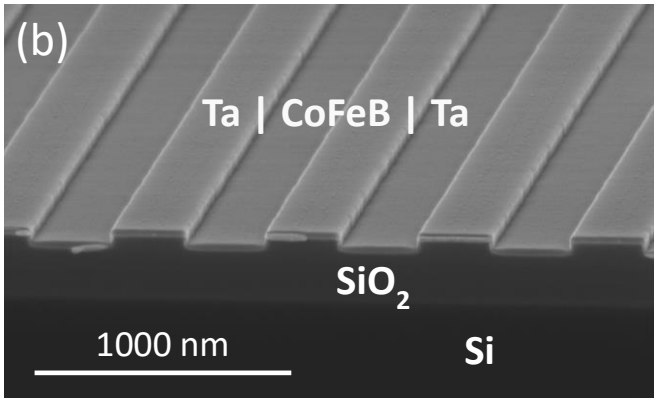
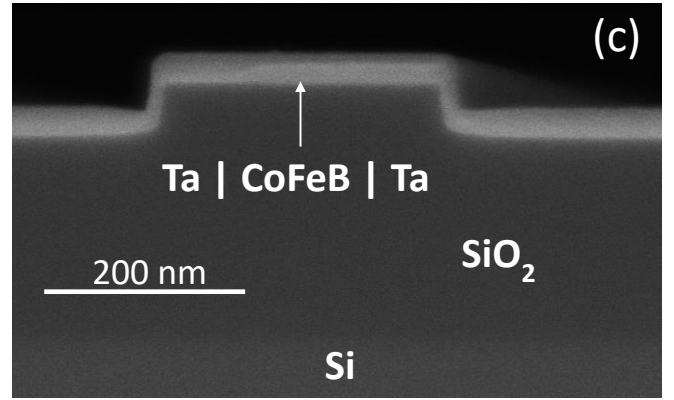

**Figure 1.** (a) Cross-sectional sketch of the meander-shaped film unit cell used in the micromagnetic simulations and in the transfer matrix calculations. The unit cell has periodicity $a=L_1+2L_2=600$ nm where $L_1$ and $L_2$ are the lengths of the horizontal segments having thickness $d_1$. The vertical segments have width $d_2$, the meander depth is $h=50$ nm and the total menader thickness is $S=h+d_1$. The dash-dot line represents the coordinate ($\xi$) along with the meander structure. An external magnetic field is applied along the z-direction and propagation occurs along the *xy*-plane. $k_1$ and $k_2$ represent the SW wavenumbers along with the horizontal and vertical segments, respectively. $\Psi_n$ and $\Psi_{n+1}$ are the SW amplitude at the input ($x=-a/2$) and output ($x=+a/2$) sections of the meander-structure unit cell. (b,c) SEM images of the meander-shaped 23-12 CoFeB film.

**Spin wave spectra measured by Brillouin light scattering spectroscopy**

Brillouin light scattering (BLS) spectra from thermally excited SWs were measured in the 180°-backscattering geometry using a (3+3)-pass tandem Fabry-Perot interferometer.[24] To map the magnonic band structure we sweep the magnon wavevectors k along the *x*-direction from zero to the edge of the fourth Brillouin zone $4\pi/a=2.08\times10^7$ rad/m. An applied static magnetic field of $m_0H_0=$ 50 mT was applied along the z-direction (parallel to the thickness step of the meander-shaped film) and ensures saturation of the CoFeB magnetization, as can be inferred from of the MOKE loops presented in Fig. S1 of the Supplemental Material.

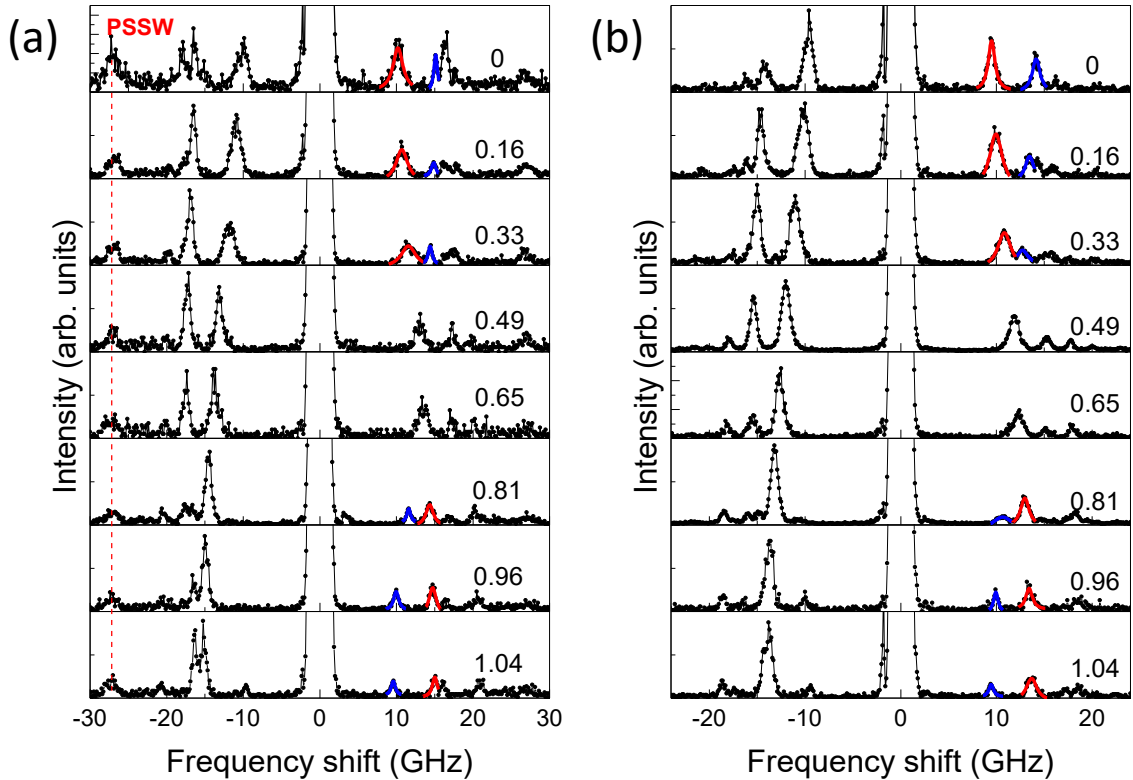

**Figure 2.** BLS spectra at $m_0H_0=50$ mT recorded at different values k values within the first two BZs for (a) 23-12 and (b) 15-8 meander-shaped CoFeB films. Labels of each spectrum correspond to the k value expressed in $10^7$ rad/m. On the anti-Stokes side of the spectra, we have highlighted the two-lowest frequency modes. For the 23-12 sample, a vertical dashed line is used as a guide for eye to indicate the position of the PSSW mode.

Figure 2 presents the sequence of the BLS spectra for the two investigated samples over the first two BZs, i.e. by varying k from 0 to $2\pi/a=1.04 \times10^7$ rad/m, where *a*=600 nm is the magnonic unit cell. Several well-resolved peaks are observed on both the Stokes and anti-Stokes side of the spectra. The red and blue highlighted peaks correspond to the two lowest-frequency SW modes which exhibit the largest frequency variation within the wave vector range investigated. The frequency separation between these two modes decreases from k=0 to k=π/$a$ where the two peaks merge and cannot be resolved anymore. On further increasing k up to k=2 π /$a$, they split again until they reach the same frequency values measured at k =0. For the 23-12 meander film, the peak at 27.2 GHz, whose frequency is independent on k, corresponds to the first perpendicular standing SW (PSSW) of the horizontal segment 23 nm thick. To note that the frequency of the PSSW mode for the 15-8 sample is well above 20 GHz and therefore the corresponding peaks are not visible in the BLS spectra.

**Micromagnetic simulations of magnonic band structure**

The magnonic band structure for the vertical meander-shaped films was calculated using micromagnetic simulations. The results are compared in Fig. 3 to the experimental dispersions. The CoFeB films were discretized into micromagnetic cells of dimensions D$x$= D$y$ = D$z$= 1×1×1 nm$^3$. A bias magnetic field of 50 mT was directed along the $z$-axis. To excite SWs in the system, a sinc-function magnetic field pulse directed along the y-axis with an amplitude of 1 mT and with a cut-off frequency of 35 GHz was applied in the excitation region of Fig. S2). The SW propagation characteristics were obtained by means of a Fourier transform of the temporal evolution of the average magnetization component along the y-direction (M$_y$($t$)) over a simulation time of 300 ns within the detection region of Fig. S2. The spatial profile of each mode in one period of meander structure was extracted by plotting M$_y$ component of dynamic magnetization for the frequencies of the SW modes through the continuous wave excitation regime of the SW signal. In Fig. 4, we plot the spatial profiles of the modes at the bottom and top frequencies of the BG$_1$ ($f_3$ and $f_4$) and BG$_2$ ($f_5$ and $f_6$). Magnetic parameters, i.e. the saturation magnetization M$_s$ = 1400 kA/m 10$^6$ A/m and the exchange constant of A$_{ex}$ = 2.1 × 10$^{-11}$J/m, are extracted from measurements of the frequency dispersion of SWs in plane films with thicknesses of 23 and 15 nm (See Fig. S2). More details on micromagnetic simulations are given in the Methods section.

**Discussion**

The frequencies of the SW modes were extracted by fitting the peaks in the BLS spectra (displayed in Fig. 2) with Lorentzian functions (red and blues curve in Fig. 2 (a) and (b)), and were plotted as a function of their respective wave vectors (k) up to the fourth BZ, as shown in Fig. 3. The magnonic band structure shows the periodic character of the branches of the dispersion realtion as well as the frequency ranges in which SWs are forbidden to propagate through the crystals. Note that the overall features of the measured dispersions are well reproduced by the micromagnetic calculation (curves in Fig. 3) from both a qualitative and a quantitative point of view.

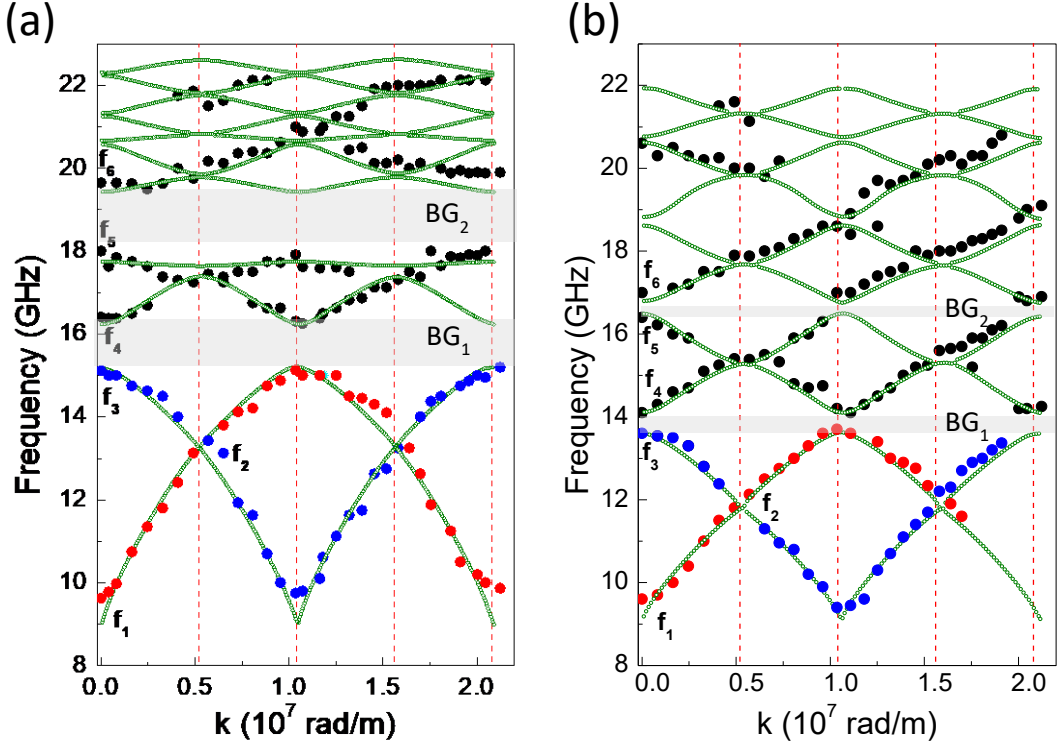

**Figure 3.** Measured magnonic band structure for the vertical meander-shaped (a) 23-12 and (b) 15-8 CoFeB films. Red and blue points represent the frequency of the lowest-two frequency peaks represented in Fig. 2 while micromagnetic dispersion data are represented by the green curves, respectively. The grey regions indicate the $BG_m$ frequency ranges for ($m$=1 and 2).

Let us first analyze the band structure for the 23-12 sample. The amplitude of the magnonic band is more pronounced for the two lowest frequency modes and decreases for the highest ones. The frequencies of this mode doublet oscillate in anti-phase within the same frequency range, i.e. from $f_1$=9.6 to $f_3$=15.1 GHz. Their group velocities $v_g = \frac{d\omega}{dk}$ at k=0 have opposite signs and differ by one order of magnitude 7.25x10$^3$ and -5.23 x10$^2$ m/s for the first and second lowest-frequency modes, respectively. The resulting magnonic bandwidth of these two modes is $f_3$-$f_1$=5.5 GHz demonstrating a rather large frequency range over which the SWs propagate along with the vertical meander structure. The third lowest frequency mode has $v_g$=2.2x10$^2$ m/s with a magnonic band width of ≈ 1.2 GHz while the fourth-lowest frequency mode has a much less pronounced dispersive character. Between each successive doublet of modes, full magnonic band gaps (BGs), where SWs are forbidden to propagate through the crystals, are observed. These are highlighted by the grey areas in Fig. 3.

The width of the first $BG_1$ was measured to be 1.05 while $BG_2$ is 1.65 GHz. To note that, for higher-order modes, the BGs are neither observed in the experiment nor predicted by the numerical calculation. Furthermore, no magnonic BG is observed between the two lowest frequency modes for $k=\pi/a=0.52\times10^7$ rad/m. Here the two lowest frequency modes (red and blues peaks of Fig. 2) merge into one single peak at frequency $f_2$ with a the non-vanishing group velocity. This is different from what was generally observed for single-[25] and multi-layer[26] 1D MCs where magnons with wavelengths satisfying the Bragg condition $k=\pi/a$ are resonantly scattered back, leading to the formation of BGs in SW spectra.

A similar frequency dispersion was measured for the 15-8 meander-shaped film, as well with an overall magnonic band shifted downward but exhibiting the same frequency value of the bottom frequency at k=0 (9.6 GHz). The two lowest frequency modes have an oscillation amplitude (4.4 GHz) between 9.6 and 14 GHz which is smaller than the one of the 23-12 sample (5.5GHz). It is important to remark that the center frequency position of the band gaps ($BG_1$ and $BG_2$) is downshifted and narrower with respect to those observed in the 23-12 sample.

Using micromagnetic simulations, we have calculated the spatial profiles of the modes at the edges of the forbidden zones $BG_1$ ($f_3$ and $f_4$) and $BG_2$ ($f_5$ and $f_6$). In Fig. 4, we display the normalized y-component of the dynamic magnetization ($M_y/M_s$) for sample 23-12. Similar results are obtained for the 15-8 sample. Since the unit cell is mirror-symmetric with respect to the vertical *y*-axis, spatial distributions for standing SWs will also have a similar type of symmetry.

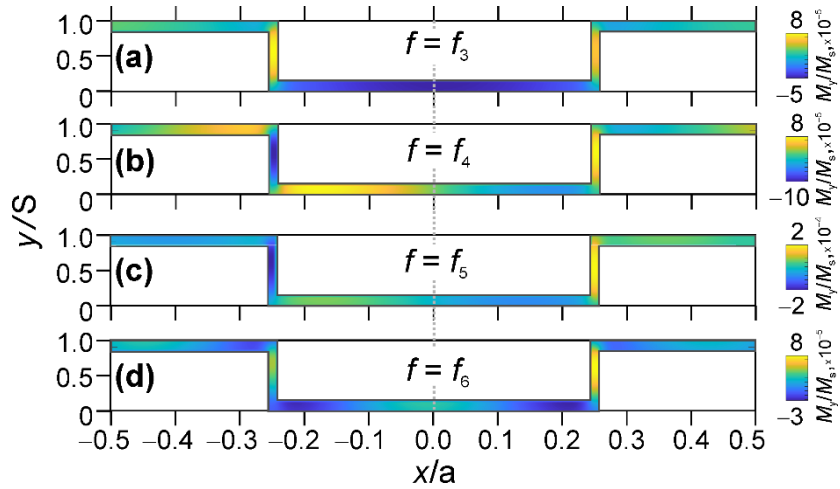

**Figure 4.** Spatial distribution of modes ($M_y/M_s$) in the meander-shaped 23-12 CoFeB film calculated by micromagnetic simulations at the edges of $BG_1$ ($f_3$ and $f_4$) and $BG_2$ ($f_5$ and $f_6$).

In particular, the modes at frequencies $f_3$ and $f_4$, which represent the edges of the first forbidden zone $BG_1$ in Fig. 3 are symmetric and anti-symmetric, respectively, with respect to the center of the meander-shaped film x=0. Contrarily, modes $f_5$ and $f_6$ for $BG_2$ are anti-symmetric and symmetric with respect to x=0. (Same mode profiles of Fig. 4 are plotted in Fig. S4 as a function of the meander-coordinate ξ). These observations suggest that the mode profiles at the BG edges correspond to standing waves with zero group velocity and the same Bloch wave number $k = \frac{2\pi m}{a}$, $m = \pm 1, \pm 2 ...$ in a periodic meander structure. These stationary waves are formed by the successive reflections of propagative waves when the Bragg condition is fulfilled. The symmetric mode profile alternates with the anti-symmetric along with monotonous frequency increase at $k = 2\pi m/a$, $m = \pm 1, \pm 2....$, while at $k = \pi n/a$, $n = \pm 1, \pm 2....$ the mode profile is subjected to the smooth transition from the top of one opened BG to the bottom of the next-order BG.

To understand in more detail the formation of the magnonic BGs in these samples, we use the transmission matrix method (TMM), which is a standard approach[14,15,27] to study the wave propagation characteristics in photonic,[28] magnonic[29,30,31] and phononic[32] crystals waveguides, to achieve a detailed analysis of the calculated frequency dispersion of all modes and their spatial profiles. The meander-shaped film is constituted by a combination of horizontal and vertical segments of regular CoFeB waveguides with thickness $d_1$ and width $d_2$, and wave vectors $k_1$ and $k_2$, respectively, as shown in Fig. 1(a). The structure periodicity is introduced by cascading of the meander-shaped periods and considering the SW amplitude at the input ($\Psi_n$ at x=-$a$/2) and output ($\Psi_{n+1}$ at x=+$a$/2) of each period as due to the interference effects between the incident, reflected and transmitted SWs. Further details of the TMM are described in the Supplemental Material.

Figure 5 (a) and (b) show the result of SW band structure calculations performed with Eq. (S10-S11) for two meander-shaped CoFeB films (23-12 and 15-8) for an external magnetic field $m_0H_0$=50 mT applied along the $z$-axis, as shown in Fig. 1 (a). The calculated band structures exhibit similar features: all modes have an oscillating frequency behavior with the same periodicity of oscillation and BGs width (Df) which decreases monotonically with increasing mode frequency ($BG_1$>$BG_2$>$BG_3$). The magnonic bandwidth and BG for the 23-12 sample are larger than for the 15-8 sample which, on the contrary, exhibits a larger number of modes in the same frequency range.

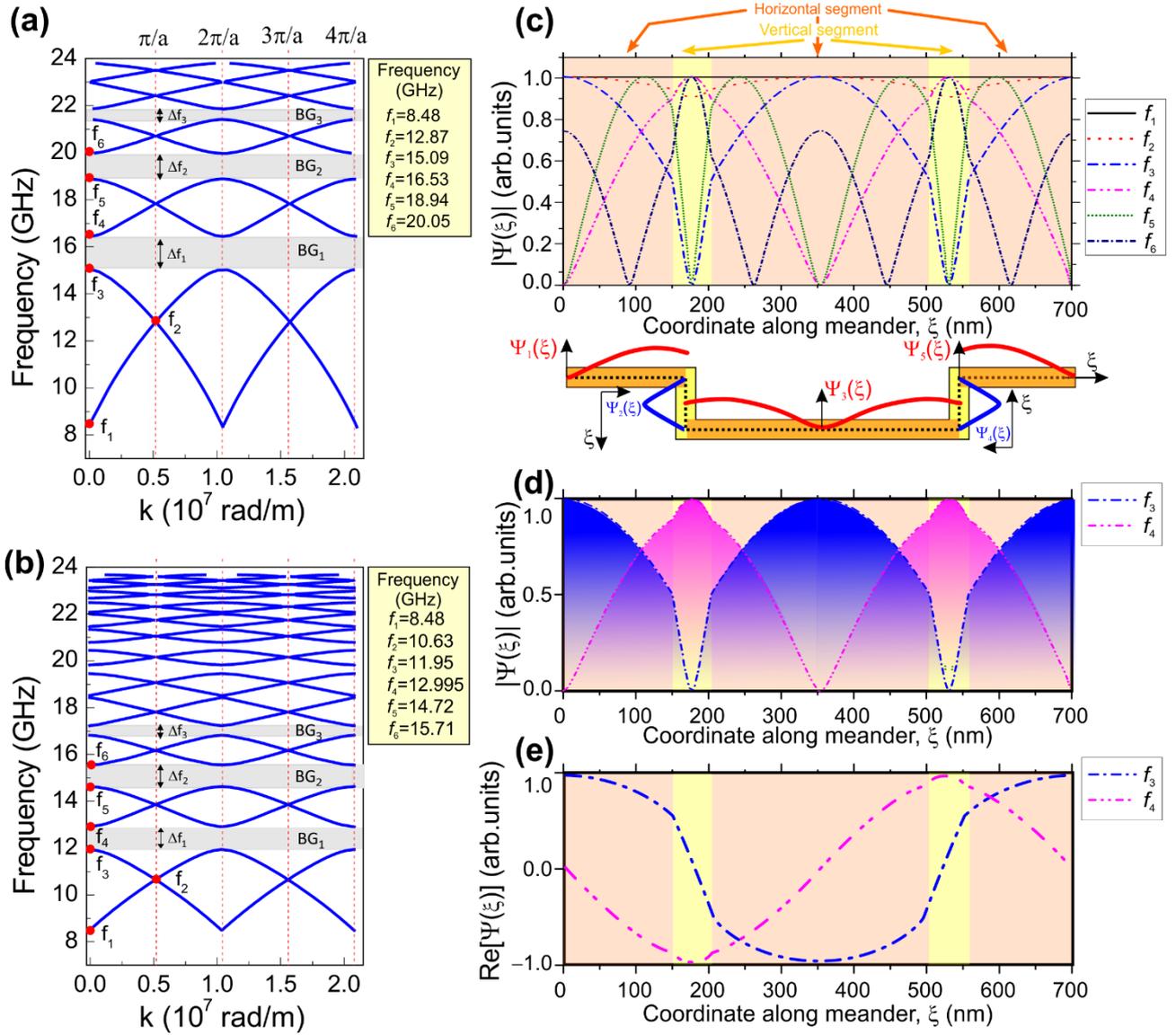

**Figure 5.** (a) Dispersion diagram of dipolar SWs in a meander-type structure with thicknesses $d_1= 23$ and $d_2 = 12$ nm. The position and the width of the first three forbidden zones ($BG_{1-3}$, $\Delta f_{1-3}$) are depicted. The dashed lines indicate the values of the Bloch wave numbers $k_m=m\pi/a$. The characteristic frequencies $f_{1-6}$ are denoted with the dots. (b) Same of (a) for $d_1= 15$ nm and $d_2 = 8$ nm. (c) Scheme for setting the eigenfunction form along the corresponding segments of the structure with coordinate $\xi$ together with the absolute value of the spatial SW mode amplitudes ($\Psi$) at different frequencies $f_{1-6}$. d) Enlarged representation of the absolute values and (e) the real part of spatial distribution of SW amplitudes at frequencies $f_3$, $f_4$ demonstrating the localization either in vertical or in horizontal segments and symmetry with respect to the center of the unit cell (x=0).

In addition, when comparing the results of Fig. 5 (a) to those calculated by micromagnetic simulations in Fig. 3 (a) for sample 23-12, we notice that the two approaches provide similar results for the dispersion of the lowest three modes while for the fourth-lowest frequency mode micromagnetic calculations provide an almost constant frequency value. Moreover, TMM predicts that the frequency width of the BGs decreases on increasing the mode frequency while this is not the case for the micromagnetic band structure of Fig. 3. These discrepancies are mainly ascribed to the fact that intralayer-exchange interaction is neglected in the TMM calculations.

The results of the TMM numerical calculations for the spatial distribution $|\Psi|$ along the coordinate $\xi$ at frequencies from $f_1$ to $f_6$ are shown in Fig. 5 (c) for the 23-12 sample. The lowest-frequency mode at frequency $f_1$ is uniformly distributed along the meander-shaped CoFeB film and resembles the Kittel mode of ferromagnetic resonance. Mode $f_2$ has a similar profile as $f_1$ but presents a shallow minimum in the vertical segment. The dispersion characteristics in the vicinity of $f_2$ have two passbands connecting at a point $2kL_1 = \pi$ and corresponding to symmetric and antisymmetric modes with respect to the longitudinal plane at x=0. At $2kL_1 > \pi$, the symmetric mode becomes antisymmetric and vice versa. To explain the absence of magnonic BG between the lowest frequency modes, it is important to remark that the dispersion characteristics of waves in periodic structures are determined not only by the material parameters of the media and the geometry of the unit cells but also by the properties of their symmetry.

The CoFeB-based meander structure studied in this article has mirror symmetry with respect to the yz-plane passing through the point $x = 0$ and a gliding plane of symmetry (glide reflection symmetry). A unit cell of a meander structure (Fig.1 (a)) can be obtained from a part of a cell of length a/2 by performing two operations: reflection relative to a plane parallel to the coordinate plane (x, z) and passing through the point y = S/2 and translation along the x axis a/2 distance. A similar type of symmetry has the Microstrip Meander-Line Slow-Wave Structure (MMLSWS) widely used in Microwave and RF Vacuum Electronic Power Sources.[29,30] In Refs. 31 and 32, the dispersion characteristics of electromagnetic waves in MMLSWS, mode symmetry, and forbidden bands were analyzed by the Green's function method in the quasi – TEM approximation. In particular, it was shown that the normal modes of MMLSWS are the even and odd modes with respect to the mirror plane of symmetry. Thus, in the frequency region where $k = \frac{n\pi}{a}$, $n = 1,3,5...$ the band gaps are absent in contrast to one-dimensional planar magnonic crystal [33]. Band gaps are observed at $k = \frac{m2\pi}{a}$, $m = 1,2,3...$, only. Consequently, the periodicity of the frequency oscillation is $4\pi/a$ and not $2\pi/a$. Similar considerations hold for the 15-8 meander-shaped film.

At the frequency $f_3$, the global minimum of the spatial distribution ($|Y(x) = 0|$) is localized in the vertical segments, while the global maxima are localized in the horizontal segments (see Fig. 5 (c)). At frequency $f_4$, the distribution is changed to the opposite. A similar variation in the spatial distribution in the vertical segments during the transition from $f_3$ to $f_4$ is also observed at the edges of the $BG_{2,3}$ zones. For example, the spatial distribution of the electromagnetic energy density at the edges of the forbidden zones in a 1D photonic crystal with modulation of the refractive index of the medium was observed in Ref. 34.

In Fig. 5 (e) we plot the amplitude of the spatial profiles Y(x) of modes corresponding to the frequencies of the bottom ($f_{b1}=f_3$) and top ($f_{t1}=f_4$) of the first BG. At the frequency $f_{b1}=f_3$, the function Y(x) is symmetric with respect to x=0 while at frequency $f_{t1}=f_4$, it is anti-symmetric, in agreement with results of micromagnetic calculations of Fig. 4.

To obtain further insight into the magnonic band structure of the meander-shaped film, we analyzed the dependence of the BG width ($\Delta f_m$) and the center frequencies $f_{cm}$ on the width $d_2$ of the magnetic vertical segment for fixed values of $d_1$=23 nm and h=50 nm, for the different modes numbers m=1,2,3. In the framework of the used TMM, the internal magnetic field in all segments of the meander structure is assumed to be uniform and the SW reflection coefficients $\Gamma = \frac{d_1 - d_2}{d_1 + d_2}$ are determined only by the ratios of the thicknesses and width of the horizontal and vertical segments, respectively (as in Eq.(2) of Supplemental III).

The first observation we derive from Fig. 6 (a) is that the $BG_m$ width is not monotonic as a function of $d_2$ in the range between 0 and 40 nm. When $d_2>d_1$, there is no significant difference between the $BG_m$ width for the different modes. When $d_2=d_1$=23nm the SW reflection coefficient in the meander structure is $\Gamma = 0$ and therefore SW propagates without reflection from the joints of segments, i.e. forbidden zones for all the modes disappear. For $d_2<d_1$, the propagation constant of SWs $k_2$ in the vertical segments increases, since for any frequency ω from dispersion equation (Eq. 12 of Supplemental) it follows that

$$k_2(\omega) = \frac{d_1}{d_2} k_1(\omega) \ , \ |\Gamma| > 0 \ .$$

and an almost linear increase in the BG width is observed. The zone with the number *m=1* has the largest ($\Delta f_m$) which decreases on increasing the band gap order *m*. On further reduction of $d_2$, the $BG_1$ width increases, reaches a maximum at $d_2$ =3.5 nm, and then vanishes for $d_2$ approaching to zero. On the contrary $BG_2$ and $BG_3$, have an oscillating behavior with minima and maxima whose position depends on *m*. This means that by choosing the width $d_2$ of the vertical segments, one can selectively control the width and center frequencies of the forbidden zones with different indices *m*. From the dependency $\Delta f_m(d_2)$ (see Fig. 6 (a)), we define the corresponding zone center frequencies $f_{cm}^j(d_{2m}^j)$ which exhibit a monotonic increase up to about 7.5 nm followed by an almost constant value for larger $d_2$ width (Fig. 6 (b)).

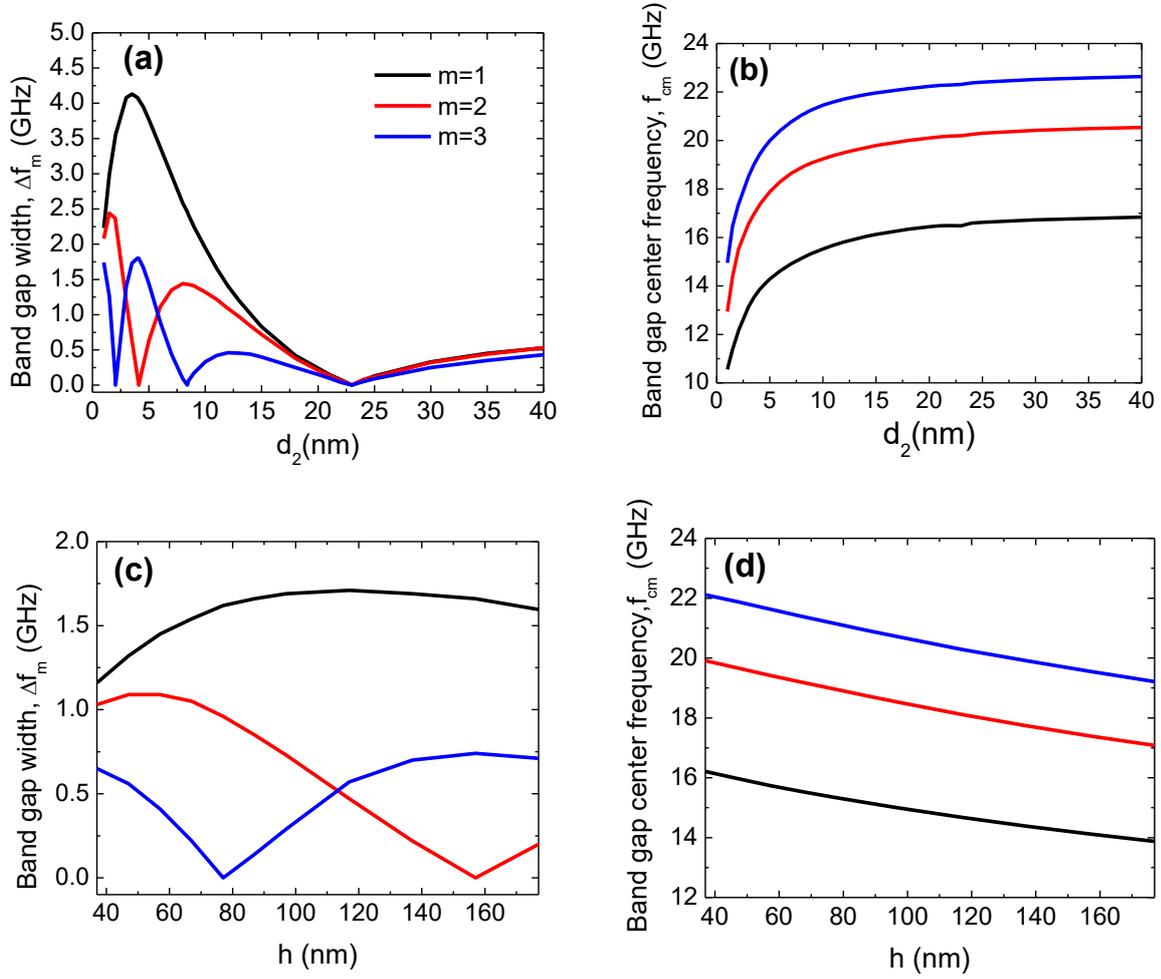

**Figure 6**. (a) Dependences of the BG width ($\Delta f_m$) and (b) center frequencies $f_{cm}$ on the width of the vertical segment $d_2$ of the meander-shaped film with $d_1$=23 nm. (c) BG width ($\Delta f_m$) and (d) center frequencies $f_{cm}$ as a function of the meander depth h with $d_1$=23 nm and $d_2$=12 nm. Curves are plotted for three lowest frequencies $BG_m$ (m=1,2, and 3).

For the meander structure with parameters $d_1$=23 nm and $d_2$=12 nm, we also calculate the dependences of the $BG_m$ width $\Delta f_m$ (Fig. 6 (c)) and the center frequencies $f_{cm}$ of the forbidden zones (Fig. 6 (d)) with numbers $m$=1,2,3 as a function of the meander depth h in the range between 37 and 177 nm. The results show that the dependencies of the width $\Delta f_m$ with an increase in the meander depth $S$ have a non-monotonous form. In particular, at certain values of $h$ the BGs dispersion characteristics of SWs vanish for $m$=2 and $m$=3. On increasing $h$, the central frequencies of the forbidden zones $f_{cm}$ are monotonously decreased. Similar considerations hold for the 15-8 sample (see. Fig. S5) even if the BG width vanished at different $d_2$ and $h$ values with respect to the 23-12 sample. Thus, by a careful choice of the meander-height $h$, one can selectively control the width and center frequencies of the forbidden zones with different indices $m$.

In conclusion, the spin-wave band structure for vertical meander-shaped CoFeB films revealed the presence of several dispersive modes with allowed and forbidden frequency gaps. This is direct evidence that spin waves propagate through the vertical meander-structure making 90° turns at the junction between horizontal and vertical segments. The investigated samples behave as three-dimensional spin-wave waveguides thus making possible vertical spin-wave transport for multilevel magnonic architecture for signal processing. Furthermore, the characteristics of the magnonic band structure can be effectively controlled by changing the geometric parameters and permits us to properly design the meander structure to engineer the magnonic band structure in a controllable manner.

**Methods**

**Sample preparation**

The CoFeB meander-shaped films were processed on Si wafers using an industrial 300 mm platform. In a first step, a 250 nm thick $SiO_2$ is formed by controlled thermal oxidation of the Si substrate. Then the $SiO_2$ is patterned to form a periodic meander-like surface using 248 nm-DUV lithography and conventional reactive-ion etching chemistry. The resulting periodic grating had a height of 50 nm and 300 nm width, and a pitch of a=600 nm. Subsequently, Ta(2nm)/$Co_{40}Fe_{40}B_{20}$/Ta(2nm) films were grown by physical vapor deposition on top of the grating. The Ta serve as seed and cap layer to prevent the oxidation of the magnetic CoFeB films.

**Magneto-optic Kerr effect and Brillouin light scattering techniques**

Magneto-optic Kerr effect (MOKE) measurements were performed in the longitudinal configuration performed at room temperature, using a photoelastic modulator operating at 50 kHz and lock-in amplification. The applied magnetic field (H), directed along the thickness step, is swept from +30 mT to – 30 mT.

Brillouin light scattering spectra from thermally excited SWs were measured in the backscattering configuration by using a 200-mW solid-state laser operating at wavelength $\lambda = 532$ nm. Light is focused on the sample by a camera objective of numerical aperture 0.24 resulting in a laser spot diameter of about 30-40 micrometers. Frequency analysis of the scattered light is performed by using a (3+3)-tandem Fabry-Pérot interferometer. Since the light scattered from SWs has a polarization rotated by 90° with respect to the incident beam, an analyzer set at extinction suppresses the signal from both elastically scattered and surface phonon-scattered light. The sample is mounted on a goniometer, which allows us to choose a specified angle of incidence of light ($\theta$), with an accuracy of 1°. Due to conservation of the in-plane momentum, by changing $\theta$ it is possible to select the magnitude of the in-plane component of the SW wave vector [k = $(4\pi/\lambda) \sin(\theta)$]. To map the SW dispersion (frequency vs k), we vary k in the range from 0 to $2 \times 10^7$ rad/m in the x-direction while the magnetic field of $\mu_0 H = 50$ mT is applied in the sample plane along the z-axis,[35] in the so-called magnetostatic surface wave (MSSW) configuration. The comparison between the measured and calculated dispersion (frequency vs wave vector) for the plane 23 nm and 15nm thick CoFeB films is shown in Fig. S3.

**Micromagnetic simulations**

For modeling the dynamic response and the magnonic band structure experimentally obtained by Brillouin light scattering spectroscopy, we performed micromagnetic simulations using the open-source GPU –accelerated MuMax3 software.[36] The computational region of the meander-shaped film is depicted in Fig. 1 (a). Along the *x* and *z* axes, periodic boundary conditions (PBC) were established while along the *y* axis, the size of the region corresponded to the height S of the studied meander structures.

For the magnonic band structure calculations, the computational structure has dimensions of 8000 nm and 300 nm along the x and z directions, respectively. To obtain the dispersion characteristics, a computational region containing 10 meander periods with a total length of 6000 nm along the *x*-axis was used (see Fig. S2). From the left and right edges, the meander was extended with homogeneous CoFeB regions of constant thickness *d* and lengths of 1000 nm. To excite and detect spin-wave excitations, we used regions with a width of 100 nm located at a distance of 800 nm from the edges of the structure. The distance between the centers of the excitation and detection regions is 6100 nm. To eliminate the reflection of spin waves at the edges of the structure, absorbing regions (ABL) of 700 nm wide along the *x* axis were introduced with an increase in the attenuation parameter according to the law $\alpha(x) = \alpha_0 (x - x_0)^2$ ($x_0$ is the coordinate of the beginning of the attenuation region) from a minimum value of 0.003 to a maximum of 1.[37]

**Transfer matrix method**

We assume a meander-shaped film which is unlimited along *z*-axis while, along *x*-axis it is characterized by a period *a* and along *y*-axis the structure has a depth S. For m$_0$H applied along the z-direction, MSSWs propagate at a frequency w in the horizontal and vertical segments of a unit cell of thickness $d_1$ and width $d_2$ since the wave numbers $k_1$ and $k_2$, respectively, are perpendicular to the direction of the externally applied field. Exchange interaction as well as magnetic anisotropy are neglected in the calculations. The general dispersion equation connecting the Bloch wave number k with the frequency ω and wave numbers $k_1$ and $k_2$ of SWs has a simple form:

$$ka = arccos(\Phi) + 2\pi m, m = 0, \pm 1, \pm 2. ... \quad (S10)$$

where

$$\Phi = \frac{\Gamma^4 \cos(\chi - \delta) + 2\Gamma^2 (1 - \cos(\delta) - \cos(\chi)) + \cos(\chi + \delta)}{(\Gamma + 1)^2 (\Gamma - 1)^2}$$

and $\delta = 2k_2(\omega)(S - d_1)$ and $\chi = 2k_1(\omega)L_1$. $\Gamma = \frac{d_1 - d_2}{d_1 + d_2}$ defines the reflection coefficient from the junction of two waveguides in the case of the wave propagating in the section with larger thickness and then travelling along the section with smaller thickness.

For MSSWs propagating in the meander-shaped film, the dispersion equation has a simple form:

$$k_{1,2}(\omega) = -\frac{1}{2d_{1,2}} ln\left(1 - \frac{4(\omega^2 - \omega_h(\omega_h + \omega_m))}{\omega_m^2}\right) \quad (S11)$$

where $\omega_h = \gamma H_0$, $\omega_m = \gamma \mu_0 M_s$ and g is the gyromagnetic ratio, $M_s$ is the saturation magnetization.

**Acknowledgements**

This work has been partially funded by IMEC's industrial affiliate program on beyond-CMOS electronics as well as by the European Union's Horizon 2020 research and innovation program within the FET-OPEN project CHIRON under grant agreement No. 801055. The numerical simulation and theoretical model for three-dimensional magnonic band structure was supported by Russian Science Foundation (Project 20-79-10191). E.N.B. acknowledges the support from RFBR (18-29-27026, 19-29-03034). S.A.N. acknowledges support by Russian Science Foundation (Project 19-19-00607) and by the Government of the Russian Federation (Agreement No. 074-02- 306 2018-286) within the laboratory Terahertz Spintronics of the Moscow Institute of Physics and Technology.

**Author contributions statement**

G. G., C.A, I.R., I.A, and F. C conceived the project

D. W., S. K., G. T., A. G., and R. C prepared samples.

G. G. performed BLS measurements.

A. S. and E. N. B. performed micromagnetic simulations and theoretical calculations of the magnonic band structure.

G.G., F. C, C. A., G. T., I. A., I. R., S. A. N, and A. S. wrote the paper. All authors discussed the results and commented on the manuscript.

**Data availability**

The data that support the findings of this study are available from the corresponding author upon request.

**Additional information**

Competing financial interests: The authors declare no competing interests.